\documentclass[twocolumn]{aastex6}
\usepackage{graphicx}
\usepackage{natbib}
\usepackage{amssymb}
\usepackage{mathrsfs}
\usepackage{amssymb}
\usepackage{amsmath}
\usepackage{cases}

\usepackage{tabularx}

\def\apjl{ApJ}%
\def\apjs{ApJS}%
\def\ao{Appl.~Opt.}%
\def\aap{A\&A}%
\slugcomment{Draft Version}

\shorttitle{Rapid Nova in NGC\,300}
\shortauthors{J. Antoniadis {\it et~al.}}

\def\nova{KSP-OT-201509a}

\begin{document}

\title{Discovery of a rapid, luminous nova in NGC\,300 by the KMTNet Supernova Program}

\author{John Antoniadis\altaffilmark{1,2}, Dae-Sik Moon\altaffilmark{3}, Yuan Qi Ni\altaffilmark{3},
Dong-Jin Kim\altaffilmark{4}, Yongseok Lee\altaffilmark{4,5}, Hilding Neilson\altaffilmark{3}
}

\altaffiltext{1}{Dunlap Institute for Astronomy \& Astrophysics, University of Toronto, 50 St. George Street, Toronto,  M5S 3H4, Canada}
\altaffiltext{2}{Max-Planck-Institut f\"{u}r Radioastronomie, Auf dem H\"{u}gel 69, 53121, Bonn, Germany}
\altaffiltext{3}{Department of Astronomy \& Astrophysics, University of Toronto, 50 St. George Street, Toronto,  M5S 3H4, Canada}
\altaffiltext{4}{Korea Astronomy and Space Science Institute, 776 Daedeokdae-ro, Yuseong-gu, Daejeon, Korea}
\altaffiltext{5}{School of Space Research, Kyung Hee University, Yongin 17104, Korea}

\begin{abstract} 

We present the discovery of a rapidly evolving \replaced{nova outburst}{transient} by the Korean Microlensing Telescope Network Supernova Program (KSP). KSP is a novel high-cadence supernova \replaced{search program}{survey} that offers deep ($\sim21.5$\,mag in $BVI$ bands)
\replaced{24-h}{nearly continuous}  wide-field monitoring for the discovery of early and/or fast optical transients.
\nova, reported here, was discovered on 2015 September 27 during the KSP commissioning run  in the direction of the nearby galaxy NGC~300, and stayed above detection limit for $\sim$ 22 days. 
We use our $BVI$ light-curves to constrain the ascent rate, $-3.7(7)$\,mag\,day$^{-1}$ in $V$, decay time scale, $t^{V}_{2}=1.7(6)$\,days, and peak absolute magnitude, $-9.65\leq M_{V}\leq -9.25$\,mag. We  \replaced{discuss}{also find} evidence for a short-lived pre-maximum halt in all bands.  
\added{The peak luminosity and lightcurve evolution make \nova~ consistent with a bright,  rapidly decaying nova outburst}. 
We  discuss constraints on the nature of the progenitor and its environment using archival HST/ACS images and  conclude  with a broad discussion on the nature of the system.

\end{abstract}

\section{Introduction}\label{sec:intro}
Over the past few years, wide-field variability surveys have significantly advanced   
our understanding of high-energy transients, from thermonuclear runaways 
and various types of supernovae (SNe) \citep[e.g.][and references therein]{ptf1},  
to $\gamma-$ray \citep{fermigrb} and fast radio bursts \citep{frbs}.

In the optical regime, contemporary experiments are typically sensitive to two types of explosive 
phenomena: ``local''  optical transients (OTs) with 
small peak luminosities ($-10 \lesssim M_{\rm b} \lesssim 
5$\,mag), such as classical and dwarf novae, and luminous 
OTs with $M_{\rm b} \lesssim -15$\,mag \citep[e.g.][ and references therein]{panstarrs1,ptf1,ptf2}. 
The intermediate luminosity regime, which is as of yet poorly explored,   
is thought to be populated by rare  cosmic explosions, such as rapid under-luminous 
SNe \citep{dsm+13}, accretion-induced collapse of white dwarfs \citep[WDs;][]{nk91,mpqt09}, 
fallback SNe \citep{dk13}, electromagnetic counterparts to compact-object mergers \citep{bfc13}, and orphan short-GRB afterglows \citep{tp02}.

Exploring this parameter space is challenging for two reasons: 
first, our limited  understanding of the underlying physical mechanisms makes it difficult to predict characteristic observational signatures and the extend to which these OTs blend with novae and SNe populations. Second, because these events are expected to be both rapid and rare, identification requires sampling of a sufficiently large volume with high temporal resolution, thereby driving the need for deep high-cadence surveys.

High-cadence experiments are additionally motivated by outstanding 
questions in long-standing astrophysical problems. For instance, our understanding of ``infant'' 
thermonuclear runaways and SNe  is limited, with questions regarding  
trigger mechanisms, shock break-out emission, ejecta masses, progenitor structure, 
and asymmetries still remaining \citep{smartt09,be08}.

Motivated by those questions, we have secured $\sim$20\% of 
 the Korea Microlensing Telescope Network \citep[KMTNet;][]{kmtnet2} observing time  
through 2020 for a dedicated survey focused on infant and/or rapidly-evolving OTs,
which we call the KMTNet Supernova Program \citep[KSP; see][]{ksp0}. 
\deleted{The KMTNet is a network of three identical 
wide-field ($2\degr \times 2\degr$ at 0\farcs4\ pixel$^{-1}$ sampling),
1.6-m optical telescopes built in Chile, South Africa, and Australia. }
\added{KMTNet is a network of three identical 1.6-m optical telescopes located at the Cerro Tololo Inter-American Observatory in Chile, the South African Astronomical Observatory, and the South Spring Observatory in Australia. Each telescope is equipped with a $2\degr \times 2\degr$ wide-field detector, comprised of four e2v CCDs which offer 0\farcs4\ pixel$^{-1}$ sampling} \citep{kmtnet2}. KSP is capable of providing deep ($\lesssim$ 21.5\,mag\footnote{\added{all magnitudes reported in this work are in the Vega system}}), high-cadence continuous monitoring in $B$, $V$ and $I$ bands \citep{ksp0}. 

In this paper we present the discovery of \nova,
a rapidly-evolving OT found towards a spiral arm of the Sculptor galaxy NGC\,300.
The transient \deleted{ ---which we classify as a nova based on the inferred peak luminosity and light-curve evolution---}
stands out for its rapid decay rate and showcases the potential of KSP for providing well-sampled multi-color light-curves of rapidly-evolving eruptions. 
The paper is structured as follows. 
In \S2, we provide a brief overview of the KSP data, and present the discovery of \nova,
alongside its multi-color evolution.
\added{In \S3 we discuss the nature of the transient based on the lightcurve characteristics and in \S4} we use HST archival images to place constraints on the progenitor of \nova\ and its environment.  
Finally, we explore the ramifications of our result and conclude with a brief discussion on the prospects of the KSP in nova search in \S5. 

\section{\nova: Discovery and light-curve}\label{sec:dl}

Between 2015 July 1 and 2016 January 10, KSP monitored a 15\,deg$^2$ area towards the Sculptor group, including a 4\,deg$^2$ field around  NGC~300 \citep[$d=1.86(7)$\,Mpc\footnote{The numbers in the parentheses are  equivalent
to the 1-$\sigma$ uncertainty at the last quoted digits.}; $M - m = 26.35$;][]{ngc300}. 
We collected  $\sim1300$ frames per $BVI$ band, with a mean cadence of $\sim$3.5~h and \added{an intra-day cadence ranging from $\sim10$ to 40~min in each filter}. 

\added{The data were processed using our custom KSP pipeline which mostly relies on public software. More specifically, after acquisition flat-fielding and correction  of the CCD cross talk, the data are automatically transferred to the KASI KMTNet data center for further processing. The data are then reduced using \texttt{SExtractor}\footnote{http://www.astromatic.net/software/sextractor} for source extraction, \texttt{SCAMP}\footnote{http://www.astromatic.net/software/scamp} for astrometry and absolute photometry and \texttt{HOTPANTS}\footnote{http://www.astro.washington.edu/users/becker/v2.0/hotpants.html} for image subtraction. The photometric calibration is based on 4 \,AAVSO All-Sky Survey standard stars  in the  field \citep[see][for details]{ksp0}. No color correction is applied.}  
We use 60-s exposures  which typically yield a limiting magnitude of  
$\sim$21.5 mag at S/N=5 under 1\farcs2 seeing for point sources.  
\added{The astrometric solution, which is derived using $\sim 10\,000$ unsaturated stars with counterparts in the second Hubble guide star catalogue} \citep{gsc2}, \added{accounts for scale, offset, rotation and distortion and is better than 0\farcs12.}
The photon-limited astrometric precision is generally better than $\sim$0\farcs5 under 1\farcs2 seeing. 

\deleted{The data were processed off-line  using  \texttt{astromatic.net} software along with  
custom IDL and python routines (see Moon et al. 2016 for details). }

\nova\ was discovered in the KSP data towards an NGC~300 spiral arm (Figure\,1) as a faint, rapidly-evolving OT. 
The source first appeared on an $I-$band image recorded on UT 270.76\footnote{All times are defined relevant to Jan\,1\,2015.0 UT} with an apparent magnitude of $m_{\rm I}$ = 20.7(3) mag, at $(\alpha,\delta)_{\rm J2000} =$(0:55:09.422,-37:42:16.5), 3\farcm373 ($\simeq 2$\,kpc) away from the centre of the galaxy.
\begin{figure*}
\begin{center}
\includegraphics[width=0.9\textwidth]{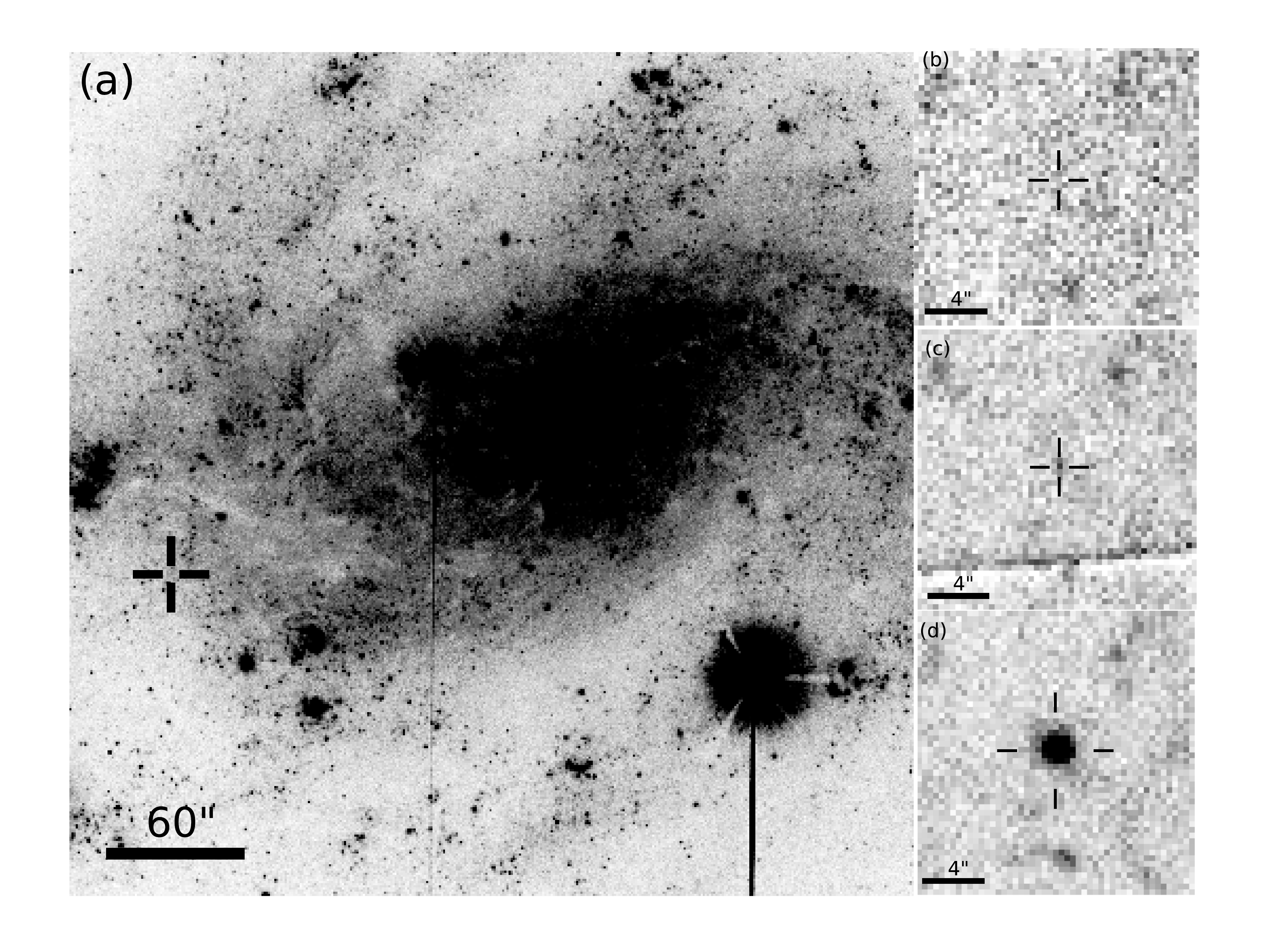}
\caption{\nova\ as recorded by KSP. Image (a) is a $B-$band image of NGC\,300, centred at its core. North is up and east is left. 
Images (b) - (d) are $I-$band images with the same orientation. (b) is an image taken  $\sim12$ minutes before detection and (c) is the first frame in which \nova~ is confidently detected. Finally (d) shows the source at its maximum $I-$band brightness}
\label{fig:lc}
\end{center}
\end{figure*}

The original light-curve produced by our automated pipeline was  contaminated by systematics, evident by the large-scale scatter ($\sim0.4$\,mag) around the dominant decay trend. 
For this reason we re-analysed the data using photometry of nearby stars. More specifically, we determined the local PSF by fitting a Moffat function plus a first degree polynomial for the background to 4 bright unsaturated stars within 3\farcm5 of the transient. 
\added{Apparent magnitudes were calculated by integrating the PSF over the Kron radius. }
\deleted{
The PSF solution was then applied to \nova, and to standard stars in the field for absolute calibration}
 Figure\,\ref{fig:lc} shows the $BVI$ light-curves of \nova\ from first detection 
to its disappearance below the detection limit $\sim$ 22 days later.

\begin{figure*}
\begin{center}
\includegraphics[width=0.8\textwidth]{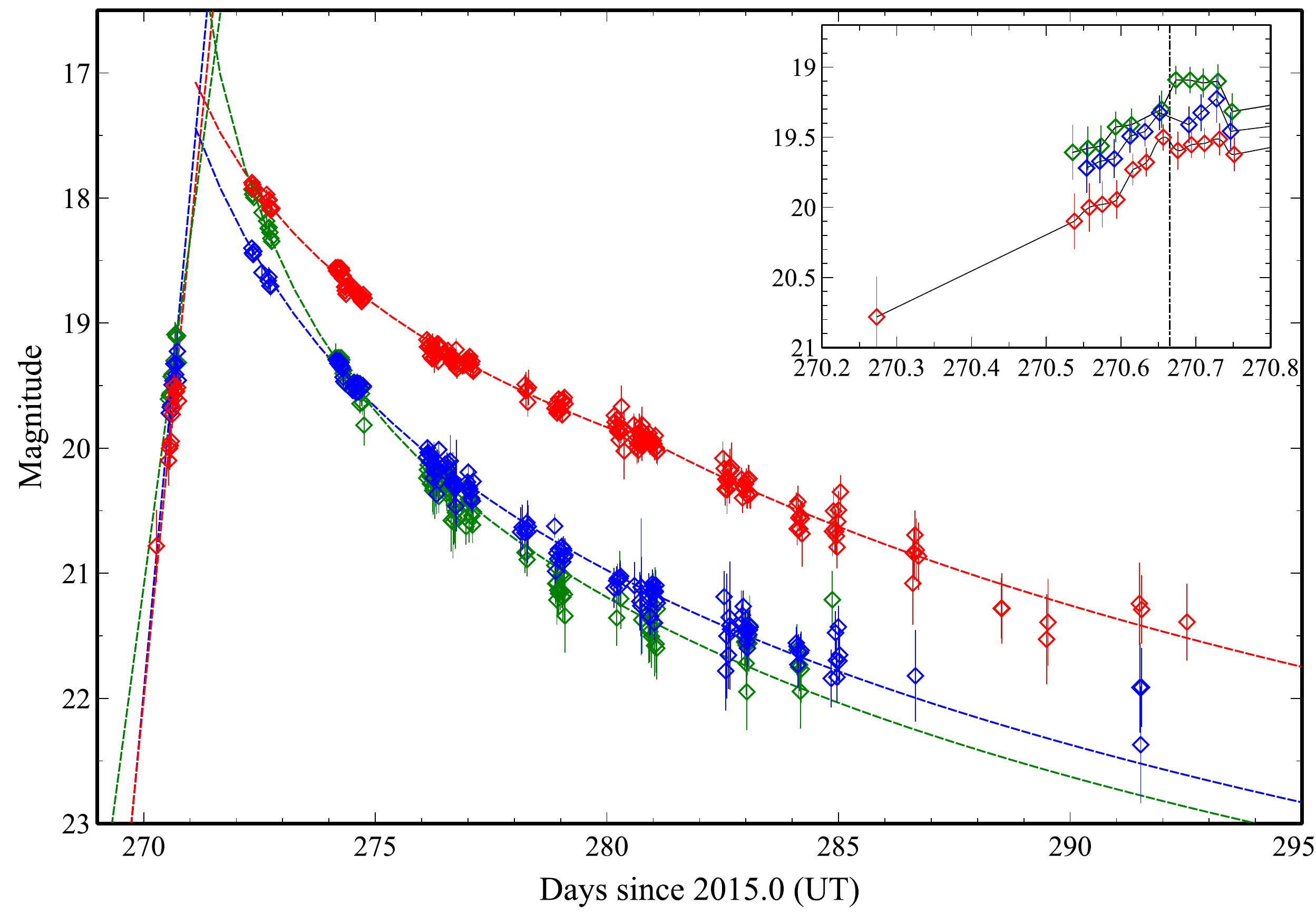}
\caption{The multi-color evolution of \nova. Blue, green and red points represent measurements in $B,V$ and $I$ respectively. The dashed lines show  the best-fit power-law decay trends ($F_{\rm v}\propto t^{-\alpha}$). The inlet figure shows the rising phase in more detail. The vertical line at UT\,270.67 indicates the approximate onset of the pre-maximum halt. }
\label{fig:lc}
\end{center}
\end{figure*}

\subsection{Rising Phase}
 
The initial phase of the transient  (Figure\,\ref{fig:lc}), sampled on 10--12 instances per each $BVI$ filter,
is characterized by a mean ascent rate of --1.9(7) mag\,day$^{-1}$ ($B$), 
--2.6(4) mag~day$^{-1}$  ($V$), and --2.5(6) mag~day$^{-1}$  ($I$), 
as determined by a linear fit to data taken prior to UT\,272. 
However, as can be seen in Figure\,\ref{fig:lc}, the ascent rate seems to decrease as the transient  progresses and remains practically constant in all bands after UT\,270.67.  

Excluding the data taken after the aforementioned time yields $-4.0(1.0)$, $-3.7(7)$ and $-3.8(9)$\,mag\,day$^{-1}$ for the mean ascent rate in $B, V$ and $I$ respectively. Based on those estimates, the probability that the halt can be attributed to random noise is $<10^{-4}$. 
Given that KSP\,N2015-09a re-appears brighter two 
nights later, we thus interpret this as evidence  for a  pre-maximum halt (PMH), often seen in light-curves of novae \citep{hpk+14}.  

\begin{figure}[h]
%\begin{center}
\includegraphics[width=0.5\textwidth]{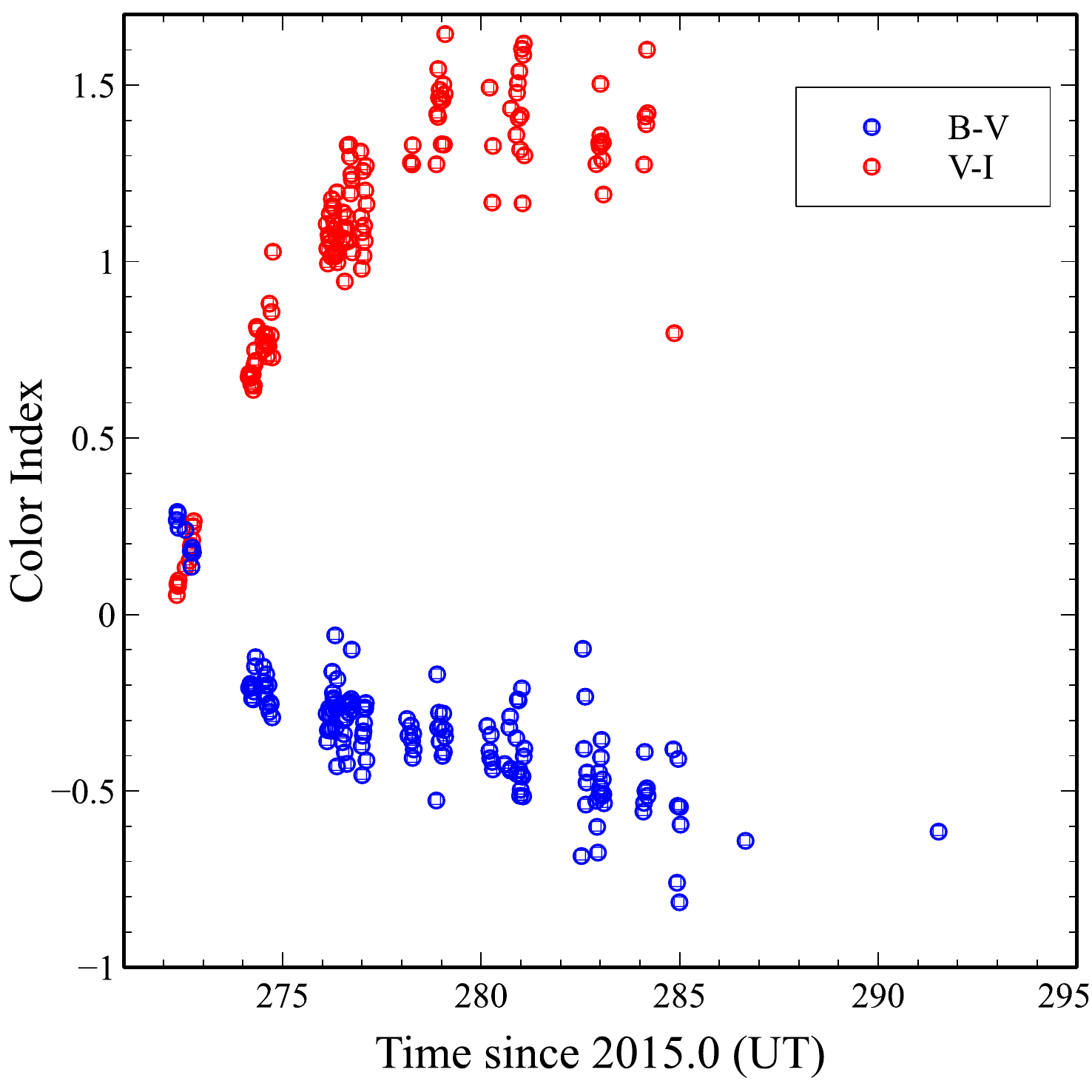}
\caption{The color evolution of \nova\ after the onset of its decay phase. 
}
\label{fig:lcm}
%\end{center}
\end{figure}

\subsection{Maximum and Decline Phase}
\nova\ was detected again on UT\,272.33 at $B$ =18.39(1), shortly after the onset of its decay phase (Figure~\ref{fig:lc}).  
An extrapolation of our best-fit models between the rising and early decline phases suggests that the peak luminosity was not missed by more than $\sim$ 0.6 days.
\deleted{Extrapolating the best-fit empirical models before and after the peak}  We place the maximum light between 17.9--17.6\,mag, 17.1--16.7\,mag, and 17.5--17.2 mag in $B$, $V$, and $I$ band, respectively.

Assuming the transient is indeed associated with NGC\,300, the former correspond to peak absolute magnitudes between  $-8.45$ and $-9.65$\,mag. We do not account for the negligible foreground extinction \citep[$A_{\rm V}^{\rm f}=0.03$\,mag;][]{sfd98}, nor for any reddening from NGC\,300, which should be of the same order since the galaxy is viewed nearly edge-on.

\deleted{The inferred luminosities are  consistent with a bright nova eruption. We adopt this classification hereafter.} 

The post-maximum evolution is characterized by a rapid decay (Figure\,\ref{fig:lc}). To quantify the 
decay rate, we adopt a decline law of the form $F_{\rm v}\propto t^{-\alpha}$, where $t$ is the time since maximum. For $B$ and $V$ we infer $\alpha=1.98(6)$ and $1.84(6)$ respectively. In the $I$ band, the decay rate evolves from $\alpha=1.41(6)$ at the onset of the decay phase  to $\alpha=2.00(7)$ after $\sim$UT\,279. 
 From the best-fit light-curves and the times of maximum-light  derived above we infer 
$t^{V}_{2}=1.7(6)$ and $t^{V}_{3}=3.8(7)$\,days for the time required for the $V-$band light-curve to fade by 2 and 3 magnitudes, respectively. 

The color evolution of \nova\ is shown in Figure\,\ref{fig:lcm}.  The decay phase starts with $B-V\simeq0.5$\,mag and $V-I \simeq -0.3$\,mag. The color indexes then rapidly evolve  to $B-V\simeq 0.0$\,mag and $V-I\simeq1.0$\,mag in less than 1 and 4 days, respectively. Subsequently,  the  excess in the $I$ band grows up to $V-I\simeq 1.5$\,mag within a few days while $B-V$ reverts to negative values.

\section{The nature of \nova}
\added{In Section\,2 we presented the lightcurve properties of  \nova. The peak absolute brightness inferred assuming association with NGC\,300 \added{and the rapid rising phase} exclude a SN explosion or a SN impostor as the origin of the transient. Similarly, the fast post-maximum evolution disfavours an outburst on a non-degenerate star and/or a luminous red nova since those generally evolve on longer timescales} \citep[cf][and references therein]{psm+07,wdb+15}.
\added{
 Those features together with the presence of a multi-color pre-maximum halt and the smooth post-maximum evolution suggest that  \nova ~ is a rapidly evolving classical nova.  Indeed, the lightcurve shown in Figure\,\ref{fig:lc} shows similarities with other well-sampled fast galactic novae such as V5589\,Sgr \citep[][see \S5 for a more detailed discussion]{ebh+17}. Finally, the observed peak luminosity/decline rate ratio is in good agreement with established empirical relations. For instance, the tangent maximum-magnitude/rate of decline (MMRD) relation of \cite{novadist} predicts $M_{\rm V}-9.0(4)$\,mag (here the error accounts for the uncertainty in $t_{2}$ and the internal scatter of the MMRD relation. For those reasons we believe \nova~ is indeed a fast classical nova. We adopt this classification hereafter.}

\section{HST Constraints on the progenitor and environment}\label{sec:hst}
We analysed a set of archival HST/ACS frames towards NGC\,300 obtained using the  f606w and f814w filters \citep[see][ for the original work]{ngc300hst}. The images were taken on 2014 July 2 with 850 and 611\,s exposure times, respectively.

We measured instrumental magnitudes and performed absolute calibration using DOLPHOT \citep{dolphot}. 
Pre-determined PSF models were used to extract instrumental magnitudes which were then corrected to infinite apertures using 12 bright isolated stars within 1\farcm5 from \nova. 
\added{Our absolute flux calibration is based on the most recent infinite apperture and zeropoint values for the ACS CCDs \citep[see][and references therein]{acscal}.}
We determine the $5\sigma$ photometric limit to be $m_{\rm f606w} \leq 27.6$ $(M_{\rm f606w}\leq 0.84)$ and $m_{\rm f814w} \leq 26.5$ $(M_{\rm f814w}\leq 0.14)$\,mag. 
We used the default ACS astrometric calibration \citep[see][]{acsastro} which provides a distortion-free system to a level of 5\,mas and then fitted for position offsets using the 4 common GSC sources nearest to \nova.

\begin{figure}[h]
\begin{center}
\includegraphics[width=0.5\textwidth]{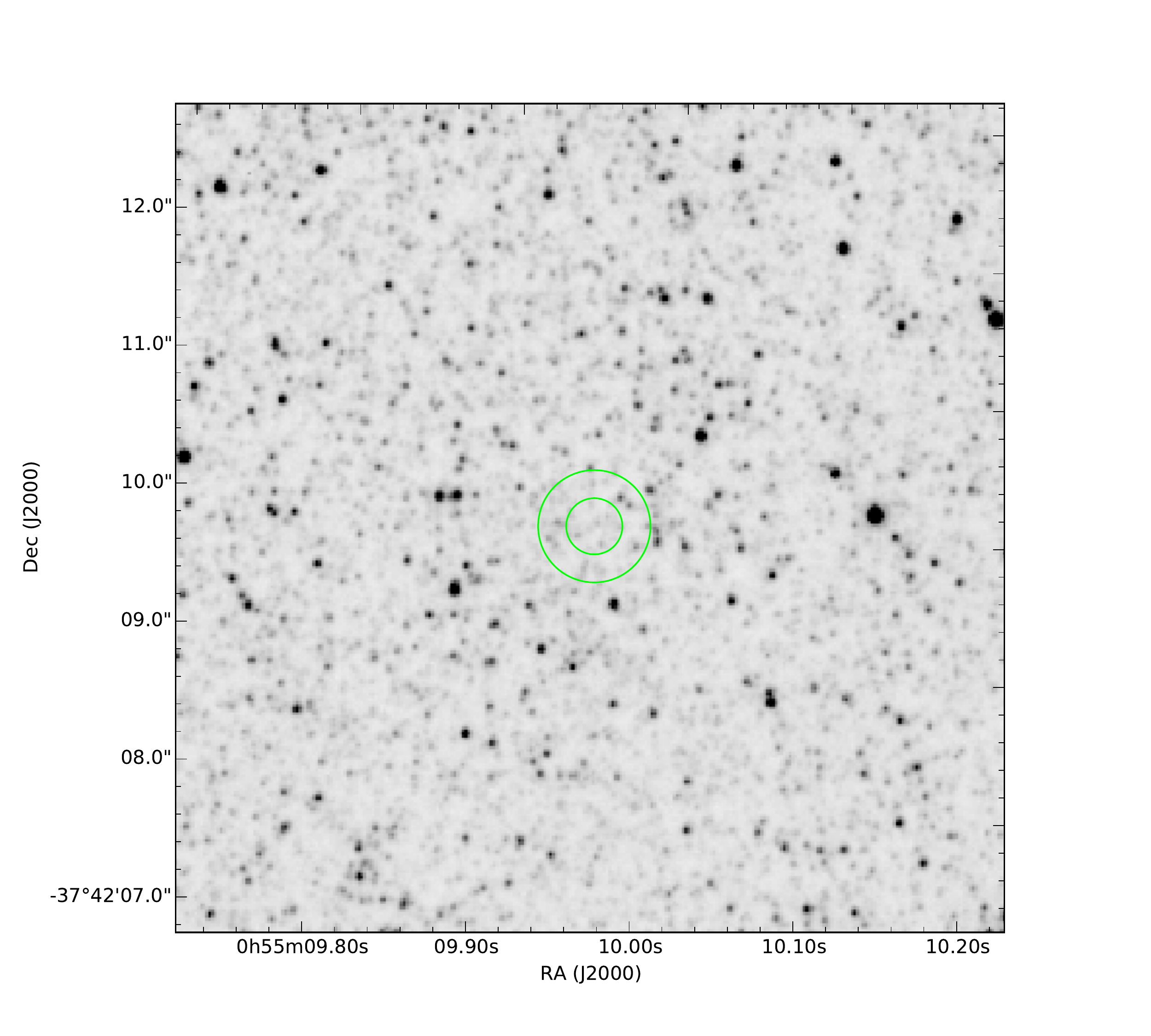}
\caption{HST image of NGC~300. The green circles, centred at the position of \nova\ represent the $1$ and $2\sigma$ astrometric uncertainties of the KSP data. The location of the brightest star is also shown.}
\label{fig:image}
\end{center}
\end{figure}

Figure~\ref{fig:image} shows the f606w image around \nova, with 0\farcs5 and 1\farcs0 error circles \added{that roughly correspond to the 68\% and 95\% KSP position uncertainty for a source close to the detection limit}. 
The star nearest to the nominal KMTNet position of the source is located at 
($\alpha$,$\delta$)$_{\rm J2000}$ = ($\rm 0^h55^m09.422^s, -37\degr42\arcsec16\farcs50$) 
and has $m_{\rm f606w} = 26.11(9)$ and $m_{\rm f814w} = 25.34(10)$\,mag. 
The brightest point-like source within the 95\% error circle has $m_{\rm f606w} = 22.348(7)$ and 
$m_{\rm f814w} = 21.344(8)$\,mag, consistent with what one would expect for a super-giant at the distance of NGC\,300. 

It is unlikely that either of the HST sources inside the KMTNet error box (Figure~\ref{fig:image}) 
is the progenitor of \nova\ in quiescence, as one would expect $ 4 \leq M  \leq 8$ for a typical main-sequence companion, which is well below the sensitivity of the HST data. 
The broader region, which is part of an NGC\,300 spiral arm, is characterized by a large number of bright stars (with $M_{\rm f606w}\simeq -5$). This indicates that \nova~ is likely associated with a region with high star-forming activity.

\section{Discussion}\label{sec:dis}
Nova outbursts result from thermonuclear runaway eruptions on a white dwarf (WD).
They occur  in a binaries of the cataclysmic-variable type, 
in which matter is accreted from a non-degenerate companion \citep[e.g.][]{drb+12}. 
Material accumulating on the WD surface eventually causes the envelope to become electron degenerate, 
leading to a runaway thermonuclear flash which ultimately gives rise to the nova phenomenon. 

It is well established, both theoretically and observationally, that nova time scales,  
amplitudes and repentance rates depend sensitively on the WD mass,  
mass-accretion rate, envelope mass, companion type and wind power \citep{yps+05}. 
With few exceptions, bright and fast novae (hereafter FNe) occur in systems with massive WDs  
and high mass-accretion rates between $\dot{M}_{\rm acc} \simeq$ 10$^{-8}$ and 10$^{-4}$ M$_{\odot}$~yr$^{-1}$ \citep{pk95,yps+05}. 
 
FNe rising phases  last up to few days \citep[e.g.][]{khc09,ssh10}. During this time, the effective  temperature increases dramatically (initially at constant radius) causing the surface brightness to rise by 10--20 magnitudes. 
Recent studies find no strong correlation between the ascent rate 
and peak brightness \citep[e.g.][]{ckn+12}, although no safe conclusions can be drawn from existing data. 
In addition, because of the rapid evolution time scales, very few infant FNe have been sampled with sufficient temporal cadence in multiple band to probe the underlying eruption mechanism in detail. 

The work presented here provides an unprecedented multi-color view of an early FN eruption phase. We find that the brightness of \nova\ increases at 
$-3.7$\,mag\,day$^{-1}$ in $V$, indicating a total rising-phase duration between $\sim2.5$ and 5 days.

In addition, our data provide evidence for a short-lived pre-maximum halt (PMH) after UT\,270.67 (Figure\,\ref{fig:lc}).
While  PMHs have been observed both in slow and fast novae \citep[see ][and references therein]{smei1,smei2}, it is yet unclear if they reflect an intrinsic change of the WD. 

For instance, some early studies suggest that they may be triggered 
by an external condition such as a sudden enhancement of mass loss from the donor. 
More recent theoretical work based on detailed 1D simulations \citep{hpk+14} finds 
that PMHs are explained naturally by a decrease in the convection-transport efficiency. 
The rise to peak brightness continues after the opacity decreases for radiation-transport to take over at the onset of the mass-loss phase. 
In \nova~ \added{one sees that the color index remains constant during the PMH and later evolves rapidly between the}  late-rise and early-decline phases (see Figures~\ref{fig:lc} and \ref{fig:lcm}). This is consistent with a transition in the emission mechanism expected in the latter scenario (see below). 

In the early decline phase
it is expected that the continuum spectrum is dominated by free-free scattering 
above the optically-thick photosphere. 
For the idealized case of pure optically-thin thermal Bremsstrahlung, 
\cite{khc09} find a universal decay law, $F_\nu$ $\propto t^{-1.75}$,
which matches well the observed lignt-curve of \nova\ (\S~\ref{sec:dl}). 
This in turn suggests that the transient evolution depends strongly on the wind rate and velocity, and less so on the WD mass \citep[][and references therein]{khc09}. 
At later phases, especially after UT 276.9, the transition to a steeper  $I$-band decline rate 
suggests the presence of an additional thermal emission component \citep{hk16} 
which may \added{indicate the presence of additional emission components, for instance} the formation of a dust shell. 
 
Considering that no spectroscopic information is available for \nova, 
we resort to other historical FNe and theoretical studies to draw further conclusions on the WD mass and the nature of the donor star as below.   
Examples of well-studied bright ($-8.5 \leq M_{V} \leq - 10.5$) FNe in the Galaxy include 
\citep{novacat} V838\,Her ($t_{2}\simeq1$\,day),  V1500\,Cyg ($t_{2}\simeq2$\,days), V2275\,Cyg ($t_{2}\simeq3$\,days) and more recently V2491\,Cyg \citep[$t_{2}\simeq 2$\,days;][]{smei2} \added{and V5589 Sgr} \citep[$t_{2}\simeq2.5$\,days][]{ebh+17}. Spectra from the early eruption phases for these FNe indicate terminal wind velocities of $\sim 1000$ to 3000 km\,s$^{-1}$. 
In almost all cases, FNe are associated with massive ($\geqslant 1$\,M$_{\odot}$) WDs and a total wind mass-loss 
of few times 10$^{-6}$\,M$_{\odot}$. 
Given the observational similarities between \nova\ and these FNe, 
we therefore conclude that the former most likely  also hosts a  massive WD of $\geqslant 1$\,M$_{\odot}$. 
\deleted{We note that the peak luminosity and decline rate also are in good agreement with  the established empirical relations. For instance, the tangent maximum-magnitude rate of decline (MMRD) 
relation by  yields $M_{V} = -9.0(4)$ (here the error includes the uncertainty in $t_{2}$ and the scatter of the MMRD relation).}

The discovery of \nova\ in an early phase and dense multi-color monitoring 
of its light-curve through its entire eruption demonstrates the potential of KSP to provide an unprecedented view of novae and related phenomena.
Based on the KSP early performance and sensitivity so far, \added{we expect to detect 
several tens} of classical novae as well as other transients, including infant and nearby SNe \citep{ksp0}. 

\acknowledgements 
We thank the anonymous referee for the useful suggestions. 
This research has made use of the KMTNet system operated by the Korea Astronomy and Space Science Institute (KASI)
and the data were obtained at three host sites of CTIO in Chile, SAAO in South Africa, and SSO in Australia.
KSP is partly supported by an NSERC Discovery Grant to DSM.
JA is a Dunlap Fellow at the Dunlap Institute for Astronomy and Astrophysics at the University of Toronto.  The Dunlap Institute is funded by an endowment established by the David Dunlap family and the University of Toronto.  
We have made extensive  use of NASA's Astrophysics Data System.
This research made use of Astropy, a community-developed core Python package for Astronomy.

\facility{KMTNet, HST(WFPC2)}
\software{Astropy}

\bibliographystyle{apj}
\bibliography{author}

\end{document}